\documentclass[pdflatex,sn-mathphys-num]{sn-jnl}% Math and Physical Sciences Numbered Reference Style
\usepackage{graphicx}%
\usepackage{multirow}%
\usepackage{amsmath,amssymb,amsfonts}%
\usepackage{amsthm}%
\usepackage{mathrsfs}%
\usepackage[title]{appendix}%
\usepackage{textcomp}%
\usepackage{manyfoot}%
\usepackage{booktabs}%
\usepackage{algorithm}%
\usepackage{algorithmicx}%
\usepackage{algpseudocode}%
\usepackage{listings}%

\usepackage{todonotes}
\usepackage{nicematrix}

\usepackage{pifont}% http://ctan.org/pkg/pifont
\newcommand{\cmark}{\ding{51}}%
\newcommand{\xmark}{\ding{55}}%

\usepackage[displaymath, mathlines,running]{}

\usepackage[T1]{fontenc}
\usepackage{inconsolata}
\theoremstyle{thmstyleone}%
\theoremstyle{thmstyletwo}%

\theoremstyle{thmstylethree}%

\usepackage{subcaption}
\usepackage[utf8]{inputenc}
\DeclareFixedFont{\ttb}{T1}{phv}{bx}{n}{9} % for bold Helvetica
\DeclareFixedFont{\ttm}{T1}{phv}{m}{n}{9}  % for normal Helvetica

\usepackage{color}
\definecolor{deepbturelue}{rgb}{0,0,0.5}
\definecolor{deepred}{rgb}{0.6,0,0}
\definecolor{deepgreen}{rgb}{0,0.5,0}

\definecolor{mbdblue}{rgb}{0,0,0.5}
\definecolor{mbdyellow}{rgb}{0.6,0,0}
\definecolor{mbdpink}{rgb}{0.94509804, 0.61960784, 0.69411765}
\definecolor{mbdpinkdark}{HTML}{E65173}

\definecolor{deepgreen}{rgb}{0,0.5,0}

\definecolor{mbddark}{rgb}{0.55294118, 0.7254902 , 0.89803922}
\definecolor{rwthblue}{rgb}{0.0 , 0.32941176, 0.62352941}

\definecolor{halfgray}{gray}{0.55}
\definecolor{ipython_frame}{RGB}{207, 207, 207}
\definecolor{ipython_bg}{RGB}{247, 247, 247}
\definecolor{ipython_red}{RGB}{186, 33, 33}
\definecolor{ipython_green}{RGB}{0, 128, 0}
\definecolor{ipython_cyan}{RGB}{64, 128, 128}
\definecolor{ipython_purple}{RGB}{170, 34, 255}

\def\code#1{\texttt{#1}}
\usepackage{listings}

\newcommand\pythonstyle{\lstset{
language=Python,
basicstyle=\ttm,
morekeywords={access,and,break,class,continue,def,del,elif,else,except,exec,finally,for,from,global,if,import,in,is,lambda,not,or,pass,print,raise,return,try,while, as},%
morekeywords=[2]{abs,all,any,basestring,bin,bool,bytearray,callable,chr,classmethod,cmp,compile,complex,delattr,dict,dir,divmod,enumerate,eval,execfile,file,filter,float,format,frozenset,getattr,globals,hasattr,hash,help,hex,id,input,int,isinstance,issubclass,iter,len,list,locals,long,map,max,memoryview,min,next,object,oct,open,ord,pow,property,range,raw_input,reduce,reload,repr,reversed,round,set,setattr,slice,sorted,staticmethod,str,sum,super,tuple,type,unichr,unicode,vars,xrange,zip,apply,buffer,coerce,intern},%
sensitive=true,%
morecomment=[l]\#,%
morestring=[b]',%
morestring=[b]",%
morestring=[s]{'''}{'''},% used for documentation text (mulitiline strings)
morestring=[s]{"""}{"""},% added by Philipp Matthias Hahn
morestring=[s]{r'}{'},% `raw' strings
morestring=[s]{r"}{"},%
morestring=[s]{r'''}{'''},%
morestring=[s]{r"""}{"""},%
morestring=[s]{u'}{'},% unicode strings
morestring=[s]{u"}{"},%
morestring=[s]{u'''}{'''},%
morestring=[s]{u"""}{"""},%
literate=
*{+}{{{\color{ipython_purple}+}}}1
{-}{{{\color{ipython_purple}-}}}1
{*}{{{\color{ipython_purple}$^\ast$}}}1
{/}{{{\color{ipython_purple}/}}}1
{^}{{{\color{ipython_purple}\^{}}}}1
{?}{{{\color{ipython_purple}?}}}1
{!}{{{\color{ipython_purple}!}}}1
{\%}{{{\color{ipython_purple}\%}}}1
{<}{{{\color{ipython_purple}<}}}1
{>}{{{\color{ipython_purple}>}}}1
{|}{{{\color{ipython_purple}|}}}1
{\&}{{{\color{ipython_purple}\&}}}1
{~}{{{\color{ipython_purple}~}}}1
{==}{{{\color{ipython_purple}==}}}2
{<=}{{{\color{ipython_purple}<=}}}2
{>=}{{{\color{ipython_purple}>=}}}2
{+=}{{{+=}}}2
{-=}{{{-=}}}2
{*=}{{{$^\ast$=}}}2
{/=}{{{/=}}}2,
literate=
{á}{{\'a}}1 {é}{{\'e}}1 {í}{{\'i}}1 {ó}{{\'o}}1 {ú}{{\'u}}1
{Á}{{\'A}}1 {É}{{\'E}}1 {Í}{{\'I}}1 {Ó}{{\'O}}1 {Ú}{{\'U}}1
{à}{{\`a}}1 {è}{{\`e}}1 {ì}{{\`i}}1 {ò}{{\`o}}1 {ù}{{\`u}}1
{À}{{\`A}}1 {È}{{\'E}}1 {Ì}{{\`I}}1 {Ò}{{\`O}}1 {Ù}{{\`U}}1
{ä}{{\"a}}1 {ë}{{\"e}}1 {ï}{{\"i}}1 {ö}{{\"o}}1 {ü}{{\"u}}1
{Ä}{{\"A}}1 {Ë}{{\"E}}1 {Ï}{{\"I}}1 {Ö}{{\"O}}1 {Ü}{{\"U}}1
{â}{{\^a}}1 {ê}{{\^e}}1 {î}{{\^i}}1 {ô}{{\^o}}1 {û}{{\^u}}1
{Â}{{\^A}}1 {Ê}{{\^E}}1 {Î}{{\^I}}1 {Ô}{{\^O}}1 {Û}{{\^U}}1
{œ}{{\oe}}1 {Œ}{{\OE}}1 {æ}{{\ae}}1 {Æ}{{\AE}}1 {ß}{{\ss}}1
{ç}{{\c c}}1 {Ç}{{\c C}}1 {ø}{{\o}}1 {å}{{\r a}}1 {Å}{{\r A}}1
{€}{{\EUR}}1 {£}{{\pounds}}1,
keywordstyle=\ttb\color{black},
emphstyle=\ttb\color{rwthblue},    % Custom highlighting style
stringstyle=\color{rwthblue},
showstringspaces=false,
numbers=left,
numberstyle=\tiny\color{halfgray},
}}

\lstnewenvironment{python}[1][]
{
\pythonstyle
\lstset{#1}
}
{}

\newcommand\pythoninline[1]{{\pythonstyle\lstinline!#1!}}

\begin{document}

\title[Zoomy: flexible modeling and simulation software for free-surface flows]{Zoomy: flexible modeling and simulation software for free-surface flows}

\author*[1]{\fnm{Ingo} \sur{Steldermann}}\email{steldermann@mbd.rwth-aachen.de}

\author*[1]{\fnm{Julia} \sur{Kowalski}}\email{kowalski@mbd.rwth-aachen.de}

\affil[1]{\orgdiv{Chair of Methods for Model-based Development in Computational Engineering}, \orgname{Faculty of Mechanical Engineering, RWTH Aachen}, \orgaddress{\street{Eilfschornsteinstraße 18}, \city{Aachen}, \postcode{52062}, \country{Germany}}}

\abstract{Free-surface flow is relevant to many researchers in water resources engineering, geohazard assessment, as well as coastal and river engineering.
Many different free-surface models have been proposed, which span modeling complexity from the hydrostatic Saint-Venant equations to the Reynolds-averaged Navier-Stokes equations.
Particularly efficient methods can be derived by depth-averaging, resulting in dimensionally reduced models. Typically, this yields hierarchies of models -- models with a variable system structure depending on the polynomial expansion of the flow variables --  that need to be analyzed and numerically solved.

This description, analysis, and simulation are challenging, and existing software solutions only cover a specific subset of models generated by these hierarchies.
We propose a new software framework to address this issue. \textit{Zoomy} allows for an efficient description, symbolic analysis, and numerical solution of depth-averaged hierarchies of free-surface flow models. \textit{Zoomy} handles a numerical discretization in one- and two-dimensional space on unstructured grids. 

With this framework, systematic evaluation of hierarchies of depth-averaged free-surface flows becomes feasible. Additionally, our open-source framework increases the accessibility of these depth-averaged systems to application engineers interested in efficient methods for free-surface flows.
}

% 4 to 6 keywords



\maketitle

\section{Introduction}\label{sec1}

% Free-surface flow is relevant to many researchers in water resources engineering \cite{water}, geohazard assessment (such as landslides \cite{landslide}, or avalanches \cite{avalanche}), as well as coastal and river engineering \cite{river}. The scientific communities have evolved the state-of-the-art on simulating free-surface flow in their respective fields, depending on physical regimes, resolution, and computing performance requirements, leading to a multitude of different methods. 
One end of the computational performance spectrum can be defined by the hydrostatic Saint-Venant equations \cite{Chow_2009} in one space dimension. It is also known as the Shallow Water equations (SWE) in one and higher dimensions. The Reynolds-Averaged Navier-Stokes equations (RANS) \cite{Pope_2000, Wilcox_2006} characterize the other limit. Compared to typical RANS simulations \cite{Castro-Orgaz_Hager_2017}, the SWE are computationally efficient. This is due to their dimensionally reduced nature, resulting from depth averaging along the vertical extent of the flow, and the assumption of a hydrostatic pressure distribution. 

In practice, the SWE turn out to be too restrictive in many application areas, in particular when the velocity profile is of interest. 
Various alternative models exist between the low-complexity SWE and higher-complexity RANS equations, some of which are listed in Table \ref{tab-models}.

% \todo[inline]{Non-hydrostatic pressure contributions or limitations in resolving vertical velocities give rise to numerous alternative formulations of depth-averaged models, which are still numerically advantageous to 3d RANS equations. A list of other published models in the field of efficient RANS-type models is listed in Table \ref{tab-models}.}

\begin{table}[h]
\caption{Shallow flow modeling overview}\label{tab-models}%
\begin{tabular}{@{}lllll@{}}
\toprule
Model & Dimension & Hierarchical & Code   & Reference \\
      &           &              & availability &      \\
\midrule
Shallow Water Equations (SWE)   & 1D,2D   & \xmark  & \cmark & \cite{Chow_2009}  \\
Shallow Water Equations (SWE)   & 3D   & \xmark  & \xmark & \cite{Blumberg_Mellor_1980}  \\
Shallow Moment Equations (SME)   & 1D, 2D   & \cmark  & \xmark & \cite{Kowalski_Torrilhon_2019, Steldermann_Torrilhon_Kowalski_2023, Koellermeier_Rominger_2020}  \\
Non-hydr. Shallow Moments (N-SME)   & 1D   & \cmark  & \xmark  & \cite{Scholz_Kowalski_Torrilhon_2024}  \\
Serre-Green-Naghdi (SGE)    & 1D, 2D   & \xmark & \cmark & \cite{Serre_1953, Green_Naghdi_1976} \\
Vertically Averaged Momentum (VAM)   & 1D, 2D   & \cmark  & \xmark & \cite{Khan_Steffler_1996a, Khan_Steffler_1996b, Cantero‐Chinchilla_Castro‐Orgaz_Khan_2018, Ghamry_2000, Escalante_Morales_De_Luna_Cantero-Chinchilla_Castro-Orgaz_2024}  \\
Shear Shallow Flow (SSF)   & 1D, 2D   & \xmark  & \xmark & \cite{Chesnokov_Liapidevskii_2011, Gavrilyuk_Ivanova_Favrie_2018}  \\
Two-Layer Non-Hydr. (L2NH)   & 1D   & \xmark  & \xmark & \cite{Escalante_Fernández-Nieto_Morales_de_Luna_Castro_2019a}  \\
Multi-Layer Non-Hydr. (LDNH)  & 1D   & \cmark  & \xmark & \cite{Fernández-Nieto_Parisot_Penel_Sainte-Marie_2018}  \\

\botrule
\end{tabular}
\end{table}

Among the many proposed models shown in Table \ref{tab-models}, commercial and open-source code accessibility is limited. Secondly, and of particular interest for this article, is the hierarchical nature of many of these models. \\ 

% \todo[inline]{maybe add a very simple intro figure to introduce the free-surface scenario and variables like the "vertical direction" mentioned in the text}

Here, "hierarchical" refers to the variable size of the equation system. This comes from the fact that depth-averaged models in Table \ref{tab-models} are generated by describing flow variables such as velocity $(u, v, w)$ or pressure $p$ in terms of a polynomial expansion with basis functions $\phi_i(z)$, depending on the vertical direction. For example $u(t,x,y,z) = \sum_{i=0}^N \alpha_i(t,x,y) \phi_i(z)$ with coefficients $\alpha_i(t,x,y)$ and basis functions $\phi_i(z)$.

Currently, there is no unified framework for analyzing and numerically solving such hierarchical models efficiently. 
To highlight some of the software design principles, we first consider the one-dimensional SME system given by

\begin{align} \label{eq:sme}
    \begin{aligned}
        & \partial_t (h) + \partial_x \left( h\alpha_0 \right) = 0 \\
        &  \partial_t  \left( h \alpha_k  \right) \langle \phi_k, \phi_k \rangle+  \partial_{x} \,  \left( \sum_{i,j=0}^{N} h \alpha_i \alpha_j \langle \phi_i \phi_j, \phi_k \rangle +  \frac{g}{2}e_z h^2 \langle 1, \phi_k \rangle \right) \\
        & \quad\quad - u_m \partial_x  \left(h \alpha_k \right) \langle \phi_k, \phi_k \rangle 
        + \sum_{i,j=1}^{N} \alpha_i\partial_x  \left(h \alpha_j \right)  \langle \phi_i \Phi_j, \phi_k' \rangle  \\
        & \quad\quad = - \left.\left( \frac{\tilde{\sigma}_{xz}}{\rho} \phi_k\right) \right\rvert_{\zeta=0}^{1} + \left\langle \frac{\tilde{\sigma}_{xz}}{\rho}, \phi_k' \right\rangle
        -  \langle 1, \phi_k \rangle g \, h \partial_x b,
    \end{aligned}
\end{align}
for $k \in \{0, ..., N\}$, where $h$ is the height of the water, $b$ the bottom topography, $\widetilde{\sigma}_{xz}$ the shear stress, and $\alpha_0 = \bar{u}$ the mean vertical velocity and $g$ the gravitational acceleration. Furthermore, let $\langle \cdot, \cdot \rangle$ denote the depth integration along the vertical direction and $\Phi_j(\zeta) = \int \phi_j(\zeta) \, d\zeta$.

An in-depth description of technical details such as the $\sigma$-coordinate transformation (a map from $z \in \left[ b, s \right]$ to $\zeta \in \left[ 0, 1 \right]$ where $s$ is the free-surface), the general derivation, and more details on the particular notation can be found in \cite{Kowalski_Torrilhon_2019}.

Equations \eqref{eq:sme} highlight most of the requirements that we identify for hierarchical free-surface flow models:

\begin{itemize}
    \item \textit{hierarchical nature}. The system \eqref{eq:sme} consists of $(N+1)$ equations.
    \item \textit{flexible basis function definitions}. The resulting system depends on the choice of basis functions $\{ \phi_i \}$ and should be exchangeable.
    \item \textit{symbolic basis integration}. The depth integration of the products of basis functions, such as $\langle \phi_i \phi_j, \phi_k\rangle$, requires symbolic integration. This is needed for important modeling tasks, such as linear stability analysis or computation of the eigenstructure of the system.
    \item \textit{material closure}. Different material closures yield different source terms for the resulting PDE system. A flexible specification of different materials is desired.
    \item \textit{non-conservative system}. The system contains non-conservative terms for any system with $N>0$. This requires a discretization using a suitable numerical method.
\end{itemize}
Other hierarchical models, such as the VAM equations, further require the treatment of a non-hydrostatic pressure distribution. 

Fast prototyping frameworks alone, such as FenicX \cite{barattaDOLFINxNextGeneration}, Firedrake \cite{rathgeberFiredrakeAutomatingFinite2017}, or DUNE-FEM \cite{dednerPythonBindingsDUNEFEM2020a}, allow flexible model definitions and are a great candidate for implementing and comparing depth-averaged models in our area of interest. However, during model development, algebraic manipulations, analytical computations of the eigenstructures, or a linear stability analysis are critical modeling tasks that cannot be tackled by these numerical frameworks alone. \\

In the following, we describe \textit{Zoomy}, a software that enables the efficient description, symbolic analysis, and numerical solution of hierarchical free-surface models. 
This code aims to bridge the gap between isolated model development in limited one-dimensional examples and the need for application engineers to test and compare these potentially attractive models on real-world test cases. 
This particularly includes the automated exploration of different models generated by model hierarchies.

Our framework covers general dimensionally reduced free-surface flow applications with equations of the form

\begin{align}
\label{eq:PDE-structure}
\begin{cases}
    & \partial_t \mathbf{Q} + \nabla \cdot \underline{\underline{F}}(\mathbf{Q}) + \underline{\underline{\underline{N}}}(\mathbf{Q}) : \nabla \mathbf{Q} = \mathbf{S}(\mathbf{Q}, \nabla \mathbf{Q}, \nabla^2 \mathbf{Q}, ..., \mathbf{P}, \nabla \mathbf{P},  \nabla^2 \mathbf{P}, ..) \\
    &  \mathbf{R}(\mathbf{Q}, \nabla \mathbf{Q} \nabla^2 \mathbf{Q}, ..., \mathbf{P}, \nabla \mathbf{P},  \nabla^2 \mathbf{P}, ...) = \mathbf{0} \quad .
\end{cases}
\end{align}
where $\mathbf{Q} \in \mathbb{R}^N$ is a vector of $N$ unknown (transport) variables, $\nabla \cdot \mathbf{F}(\mathbf{Q})$ are conservative fluxes, $\underline{\underline{\underline{N}}}(\mathbf{Q}) : \nabla \mathbf{Q}$ is a nonconservative product and $\mathbf{S}(\mathbf{Q}, \nabla \mathbf{Q} \nabla^2 \mathbf{Q}, ...)$ is a source term. $\mathbf{P} \in \mathbb{R}^M$ is a vector of $M$ unknowns typically associated with Poisson-type equations resulting from non-hydrostatic pressure constraints defined by the residual vector function $\mathbf{R}$.
Note that all PDEs of the models stated in Table \ref{tab-models} follow the PDE structure \eqref{eq:PDE-structure}.

\begin{figure}[!htb]
    \centering
    \includegraphics[width=1.0\linewidth]{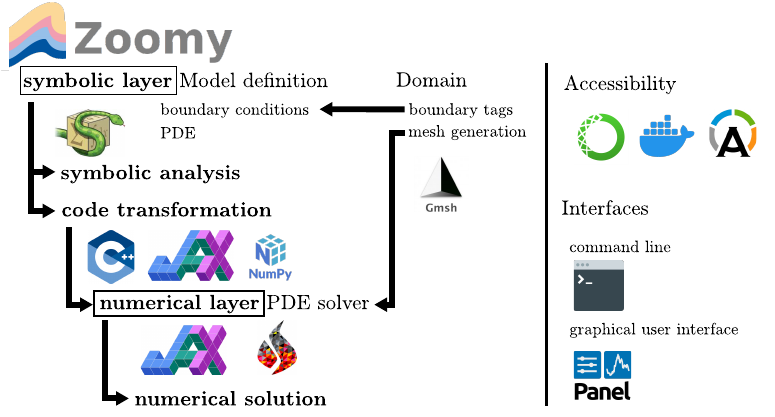}
    \caption{Zoomy software overview. The software consists of two layers, the symbolic layer and the numerical layer. The former allows symbolic mathematical operations on the PDE, while the latter generates a PDE solution numerically. The icons indicate the dependencies and usage of existing software packages in this project. Icon explanation: symbolic layer: \code{sympy}, code transformation: \code{C++, JAX, NumPy}, numerical layer: \code{JAX, FenicsX}, Domain: \code{GMSH}, Accessability: \code{Conda, Docker, Apptainer}, Interfaces: \code{Command line, Panel}}
    \label{fig:Zoomy-overview}
\end{figure}

Figure \ref{fig:Zoomy-overview} shows an overview of the general software architecture:

\begin{itemize}
    \item \textit{Zoomy} comprises a \textit{symbolic} and a \textit{numerical} layer. 
In the \textit{symbolic} layer, the Python package \code{SymPy} is used to create a symbolic representation of the PDE system \eqref{eq:PDE-structure} and the boundary conditions. 
    \item The \textit{symbolic} and \textit{numerical} layers are connected through the \textit{ code printer} available in \code{SymPy}. 
The printer allows for the conversion of symbolic expressions for different back-ends, including Python packages \code{numpy} and \code{jax} or standard \code{C}. 
    \item The \textit{numerical} layer consists of the mesh and a solver. To support unstructured meshes, we utilize meshes created with the open-source tool GMSH \cite{gmsh}. We provide components that can be used to construct customized finite volume type (FVM) solvers. Some generic implementations, such as a hyperbolic FVM solver for non-conservative system on unstructured grids or a corrector-predictor scheme for non-hydrostatic problems are available.
\end{itemize}
The code printer allows for an easy transition from the \textit{symbolic layer} to custom numerical solvers or existing discretization frameworks. In particular, we provide an interface to the Unified Programming Language (UFL) based framework FenicsX.
The derivation of depth-averaged free-surface flow models is tedious and challenging to comprehend, as the closed PDE systems comprise numerous equations and the connections to the underlying Navier-Stokes system are not readily apparent. Therefore, we see it as a critical component of \textit{Zoomy} software to be accessible and the results to be comparable. 
Accessibility is ensured by implementing \textit{Zoomy} as open-source software, with ready-to-use installations distributed as \code{conda} environments, \code{Docker} containers, and \code{Apptainer} images for high-performance computing environments. Additionally, we provide a simple GUI that can be deployed as a cloud service to promote installation-free access.
Comparability between models is gained in a postprocessing step, where the internal variables of the system $(\mathbf{Q}, \mathbf{P})$ are interpolated in the non-reduced space $(t, x, y, z)$, resulting in output variables $h, (\mathbf{u}, p)$. This relatively simple addition is crucial for interpretability, comparability of models, and a user-friendly starting point. 

The following presentation is organized as follows: In Section \ref{sec:design}, we discuss the requirements of our software, its architecture, and its features. In Section \ref{sec:examples}, we present a series of examples to demonstrate the applicability of our software to the development of depth-averaged free-surface flow solvers. In Section \ref{sec:discussion} we discuss limitations, open questions, and future directions.

\section{Software design}\label{sec:design}

\textit{Zoomy} emerged from the need to have a PDE solver capable of handling the hierarchy of models generated in the \textit{shallow moment equations} \cite{Kowalski_Torrilhon_2019}. A systematic comparison between such PDE systems of different orders would otherwise result in the manual implementation of each model, which quickly becomes infeasible for high-order systems.
Similarly, the symbolic analysis of high-order equations becomes increasingly challenging.
Additional tools, such as a graphical user interface (GUI) and scripts for automating post-processing tasks, as well as alternative PDE solver backends, are developed around this core application to make the software more convenient for free-surface flow modelers.

\subsection{Core}

The core of \textit{Zoomy} consists of a \textit{symbolic} and a \textit{numerical} layer. The key design idea is to find a suitable representation of the PDE in \eqref{eq:PDE-structure} in a symbolic form, usable for typical analytical modeling tasks, which can then be forwarded to the \textit{numerical} layer and transformed into numerical code used in the numerical solver.

\subsubsection{Symbolic layer}

The \textit{symbolic layer}  requires

\begin{itemize}
    \item the generation of PDE models from a modeling hierarchy, requiring analytical integrations of products of basis functions and a flexible size of the system in \eqref{eq:PDE-structure}, 
    \item an analytical computation of the Jacobian $\frac{\partial \underline{\underline{F}}}{\partial \mathbf{Q}}$ to symbolicly compute the eigenstructure of system \eqref{eq:PDE-structure} and allow for the transition of \eqref{eq:PDE-structure} into a fully quasilinear form for some numerical solvers,
    \item the analytical computation of the Jacobian $\frac{\partial \mathbf{S}}{\partial \mathbf{Q}}$ to analyse the stiffness of the source term or allow for the construction of implicit solvers for the source term via operator splitting,
    \item computing the linear stability analysis for non-hydrostatic formulations and
    \item the conversion of the symbolic representation into a format that is usable for computing a numerical solution.
\end{itemize}

\code{SymPy} \cite{simpy} is a Python package for symbolic mathematics and offers a \textit{code printer} to expose the formulation to the \textit{numerical layer}. 
The blueprint for a PDE model is defined in the (virtual) class \code{Model}. The part of the class specification reads as follows:

\begin{python}
class Model:
    """
    Generic (virtual) model implementation.
    """

    # User inputs
    name: str
    dimension: int
    boundary_conditions: BoundaryConditions
    initial_conditions: InitialConditions
    n_fields: int
    variables: IterableNamespace
    aux_variables: IterableNamespace
    parameters: IterableNamespace

    
    def flux(self):
        return [ZeroMatrix(self.n_fields, 1) 
                for d in range(self.dimension)]

    def nonconservative_matrix(self)
    
        return [ZeroMatrix(self.n_fields, self.n_fields) 
                for d in range(self.dimension)]

    def source(self):
        return ZeroMatrix(self.n_fields, 1)

    def residual(self):
        return ZeroMatrix(self.n_fields, 1)
\end{python}

The above code snippet highlights that the model is closely connected to the PDE, as its specification requires a symbolic representation of variables ($\mathbf{Q}$), auxiliary variables, and parameters. 
Auxiliary variables are introduced to represent constant scalar parameter fields as well as to register runtime-dependent fields such as gradients, e.g. $\partial_x \mathbf{Q_i}$ or equations of state. The update of the auxiliary variables is part of the \textit{numerical layer} described later. The class \code{BoundaryConditions} enables a symbolic representation of the boundary conditions discussed later. The definitions of the functions \code{flux}, \code{nonconservative\_matrix} and \code{source} closely match the function in \eqref{eq:PDE-structure} $\underline{\underline{F}}$, $\underline{\underline{\underline{N}}}$ and $\mathbf{S}$.

Note that in practice, there are several ways to represent \eqref{eq:PDE-structure} in the numerical implementation. For hydrostatic systems, the pressure-related variables $\mathbf{P}$ vanish, and the resulting system is typically purely hyperbolic. For non-hydrostatic systems, \textit{predictor-corrector} can be used to iteratively obtain a solution for $\mathbf{Q}$ and $\mathbf{P}$. 
Both can be represented within our framework using two models: the first model describes the underlying hyperbolic system, and a second model uses the \code{residual} function to represent the pressure-related constraint.

By inheriting from the base class \code{Model}, PDE systems such as the \textit{Shallow Water Equations} (SWE), the \textit{Shallow Moment Equations} (SME), the \textit{Vertically Averaged Momentum equation} (VAM), or the \textit{Serre-Green-Nagdhi} equations can be constructed.

Due to the symbolic representation, the analytical eigenstructure of the quasilinear matrix $\underline{\underline{\underline{A}}} \cdot \mathbf{n} = \left( \frac{\partial \underline{\underline{F}}}{\partial \mathbf{Q}} \right) + \underline{\underline{\underline{N}}} ) \cdot \mathbf{n}$ with $\mathbf{n}$ a symbolic directional vector can be computed with \code{SymPy}.

Initial and boundary conditions are necessary to define well-posed transient PDEs. Both functions can be provided to the model. A single \code{BoundaryCondition} encodes a function for a physical part of the domain using the symbolic representation of \code{SymPy}. As boundary conditions usually need to be customized for the particular PDE, we only provide a couple of generic boundary conditions, including \textit{Periodic}, \textit{Extrapolation}, and \textit{Lambda} boundary conditions. The latter provides a simple interface to define, for instance, inflow or outflow boundary conditions, as it overwrites the \textit{Extrapolation} condition for specific variables with a custom function. The following code snippet shows how we can provide a subcritical inflow boundary at the inflow, an extrapolation boundary condition at the outflow, and periodic boundary conditions on the top and bottom of the domain for the  \textit{shallow water equations} with variables $\mathbf{Q} = (h, hu, hv)^T$, where the \code{physical\_tag} corresponds to the physical name created in the \textit{GMSH} \cite{gmsh} input mesh:

\begin{minipage}[h]{1.\linewidth}
\begin{python}
import library.model.boundary_conditions as BC
boundary_conditions = BC.BoundaryConditions(
    [
        # inflow discharge of 0.1 m^3/s
        BC.Lambda(physical_tag='inflow', prescribe_fields={
            1: lambda t, x, dx, q, qaux, parameters, normal: 0.1,
        }),
        BC.Extrapolation(physical_tag='outflow'),
        BC.Periodic(physical_tag='top', 
                    periodic_to_physical_tag='bottom'),
        BC.Periodic(physical_tag='bottom', 
                    periodic_to_physical_tag='top'),
    ]
)
\end{python}
\end{minipage}

Note that no mesh needs to be defined during the initial model creation. 
Consistency of the \code{physical\_tag}s of the mesh and the model is verified at the numerical layer.

The last \textit{symbolic layer} ingredient is the representation of the model hierarchy, as presented in the introduction.
For the particular case of the SME as presented in \eqref{eq:sme}, we add two new member variables \code{level} and \code{basisfunctions}. The former is an integer defining the polynomial order of the horizontal velocity field $u$. The latter is a class tailored for the SME model, computing the analytical integrals of products of basis functions $\phi_i(\zeta)$ needed in the model formulation.
Typically, Legendre polynomials are used as basis functions. The SME of arbitrary polynomial order can then be built based on such tensors containing the integrals of products of basis functions.

For example, the SME \code{nonconservative\_matrix} of the System \eqref{eq:sme} can be written as

\begin{minipage}[h]{1.\linewidth}
\begin{python}

class Basismatrices:
    def __init__(self, basis=Legendre_shifted(), 
                 use_cache=True, cache_path=".cache"):
        self.basisfunctions = basis
        
    def _B(self, k, i, j):
    """ 
    Compute <(phi')_k, phi_j, int(phi)_j>
    """
        return integrate(
            diff(self.basisfunctions.eval(k, z), z)
            * integrate(self.basisfunctions.eval(j, z), z)
            * self.basisfunctions.eval(i, z),
            (z, 0, 1),
        )


    
class ShallowMoments(Model):
    def nonconservative_matrix(self):
        nc_x = Matrix([[0 for i in range(self.n_fields)] 
                      for j in range(self.n_fields)])
        h = self.variables[0]
        ha = self.variables[1 : 1 + self.levels + 1]
        p = self.parameters
        um = ha[0] / h
        for k in range(self.levels + 1):
            nc_x[k + 1, k + 1] -= um
            for i in range(1, self.levels + 1):
                for j in range(1, self.levels + 1):
                    nc_x[k + 1, i + 1] += ( ha[j] / h
                        * self.basismatrices.B[k, i, j]
                        / self.basismatrices.M[k, k]
                    )
        return [nc_x]
\end{python}
\end{minipage}

The code snippet shows how closely the implementation follows the mathematical formulation of \eqref{eq:sme}.

The symbolic formulation can be converted to various formats for the \textit{numerical layer}. Transformations to \code{jax}, \code{numpy}, and \code{C} are interesting for our application. The transformation requires that the interface between the \textit{symbolic layer} and \textit{numerical layer} is fixed. This means that all PDE-related functions in the class \code{Model} have a clearly defined input and output structure. 

For example, we impose that the functions \code{flux} or \code{source} depend on the three inputs \code{(Q, Qaux, parameters)}, where each input will be of type \code{jax.numpy.ndarray}, \code{numpy.ndarray} or \code{double*}, depending on the code transformation. 
Other functions depend on more inputs; e.g., the \code{eigenvalues(Q, Qaux, parameters, normals)} require an additional input for the normal direction.

A subtle yet critical problem arises when code transformations are required for vectorized expressions in \code{numpy} or \code{jax}. Using vectorized expressions is usually recommended for performance reasons, especially when using \code{numpy}. However, code transformations of constant numbers, often appearing in boundary conditions (fixed Dirichlet value) or source terms (constant source), result in non-vectorized code, as a constant number is not an array of the same length as \code{Q}. To address this issue, we employ a simple fix by adding a very small number ($10^{-20} \mathbf{Q}_0$) to each expression that contains only constants, resolving this transformation issue.

To create a run-time model in Python, the routine \code{\_create\_runtime\_model} exists, while the transformation to a C library can be issued with the routine \code{create\_runtime\_model\_C\_library}.

\begin{figure}
    \centering
    \includegraphics[width=1\linewidth]{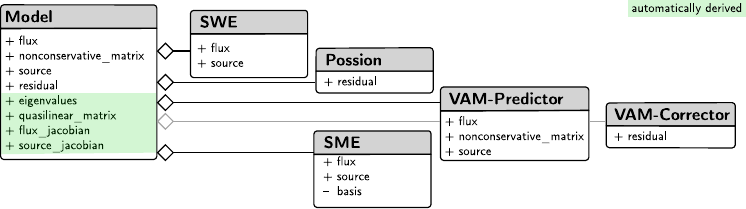}
    \caption{Classes derived from the \code{Model} base class.}
    \label{fig:models}
\end{figure}

The inheritance and class composition for the various models discussed in this paper are shown in Figure \ref{fig:models}. The inheritance of the base class \code{Model} allows the creation of new symbolic models, while the fixed interface ensures that the numerical routines remain compatible.

\subsubsection{Numerical Layer}

Numerical solvers are often tailored for the particular PDE model at hand. We provide building blocks that allow for the construction of new numerical schemes. From a top-level view, there exists a generic base class \code{Solver}, equipped with the interface

\begin{python}
class Solver:
    def _initialize(self, model, mesh, settings):
    """
    Initializes Q, Qaux, and collects the list of 
    boundary conditions based on the mesh
    """
        ...
        return Q, Qaux
    def _load_runtime_model(self, model):
        ...
        return runtime_pde, runtime_bcs
    def _get_boundary_operator(self, mesh, runtime_bcs):
        ...
        @jax.jit
        def boundary_operator(time, Q, Qaux, parameters):
            ...
            return Q
        return boundary_operator
    def solve(self, mesh, model, settings):
        ...
        return Q, Qaux
\end{python}

The function \code{\_initialize} creates the link between the symbolic model and the mesh by allocating memory for the two arrays \code{Q} and \code{Qaux}, and creating necessary maps to identify which degrees of freedom of the two vectors belong to which boundary. The function \code{\_load\_runtime\_model} triggers the transformation from the symbolic model to the numerical model with functions compatible with \code{jax}.
Boundary conditions are necessary for any system described in \eqref{eq:PDE-structure} and are therefore included in the base solver. 
Many functions, such as the \code{get\_boundary\_operator} function, follow the pattern

\begin{python}
    def get_func(self, *static_arguments):
        ...
        @jax.jit
        def func(*dynamic_arguments):
            ...
            return
        return func
\end{python}

where an outer function \code{get\_func} takes input arguments that are considered static at runtime. An inner function is built and compiled using (\code{jax.jit}) and only depends on dynamic arguments such as \code{(Q, Qaux, parameters)}.
This pattern makes it easy to track which variables will be considered for the automatic differentiation of the code. The same pattern reappears for all time-consuming building blocks that necessitate compilation.

The function \code{solve} is the public interface of the \code{Solver} class. It contains the implementation of the numerical scheme. We already provide a couple of specializations of the \code{Solver} class e.g. a transient hyperbolic PDE solver (\code{HyperbolicSolver} or a steady state solver (\code{SteadyResidualSolver}).
Figure \ref{fig:building_blocks} shows a summary of the building blocks and their usage in some solvers:

\begin{figure}
    \centering
    \includegraphics[width=1\linewidth]{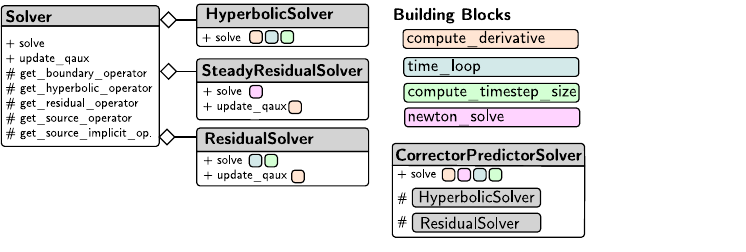}
    \caption{Building blocks and available solver classes}
    \label{fig:building_blocks}
\end{figure}
\begin{itemize}
\item The implementation of a finite-volume scheme for nonconservative systems on unstructured grids for 1D, 2D, and 3D is based on \cite{Dumbser_Enaux_Toro_2008, Castro_Fernández-Nieto_Ferreiro_García-Rodríguez_Madroñal_2009, Castro_Pardo_Parés_Toro_2009} to solve the \code{hyperbolic\_step}. 
\item Combining the \code{hyperbolic\_step} with a \code{boundary\_operator} and a time loop essentially composes a transient \textit{hyperbolic solver} (\code{HyperbolicSolver}).
\item Similarly, we implement a Newton solver (\code{NewtonSolver}) based on the GMRES linear solver available in \code{jax}. This building block can be used to create a \textit{steady residual solver} (\code{SteadyResidualSolver}).
\end{itemize}
As our code enables automatic differentiation through \code{Jax}, the residual function can be differentiated to construct the analytical Jacobian required for the Newton solver, eliminating the need for additional user input. \\

Another building block is the gradient reconstruction based on a least-squares reconstruction of arbitrary polynomial order (\code{compute\_derivatives}). The implementation handles unstructured grids in 1D, 2D and 3D. To avoid a complicated management of ghost cells at the boundaries, we restrict our implementation to a single layer of ghost cells and reconstruct gradients at the boundary by collecting a sufficiently large neighborhood of cells in the inner domain of the mesh.

Non-hydrostatic depth-averaged free-surface models require building more advanced numerical solvers.
% \todo[inline]{predictor corrector scheme lacks a bit of context for me at this point, maybe add 2 sentences of explanation}
Following an operator splitting scheme \cite{Escalante_Morales_De_Luna_Cantero-Chinchilla_Castro-Orgaz_2024}, we identify the following semi-implicit schemes as one interesting candidate to implement:
\begin{subequations}
\begin{align}
&\textit{explicit hydrostatic predictor (hyperbolic PDE)} \notag \\
& \frac{\left(\mathbf{Q}^{*} - \mathbf{Q}^{n}\right)}{dt} + \nabla \cdot \underline{\underline{F}}(\mathbf{Q}^n) + \underline{\underline{\underline{N}}}(\mathbf{Q}^n) : \nabla \mathbf{Q}^n = \mathbf{0} \label{eq:PDE-hyperbolic} \\
&\textit{semi-implicit non-hydrostatic corrector (Possion type PDE)} \notag \\
&\mathbf{R}(\mathbf{Q}^{**}, \nabla \mathbf{Q}^{**}, \nabla^2 \mathbf{Q}^{**},..., \mathbf{P}, \nabla \mathbf{P},  \nabla^2 \mathbf{P}, ...) = \mathbf{0} \label{eq:PDE-non-hydrostatic} \\
&\frac{\left(\mathbf{Q}^{**} - \mathbf{Q}^{*}\right)}{dt} = \mathbf{S}^P\left(\mathbf{Q}^*, \nabla \mathbf{Q}^*,  \nabla^2 \mathbf{Q}^*, ..., \mathbf{P}, \nabla \mathbf{P},  \nabla^2 \mathbf{P}, ...\right)\label{eq:PDE-non-hydrostatic-update} \\
&\textit{explicit (or implicit) friction term (ODE)} \notag \\
&\frac{\left(\mathbf{Q}^{n+1} - \mathbf{Q}^{**}\right)}{dt} = \mathbf{S}^F(\mathbf{Q}^{**}, \nabla \mathbf{Q}^{**}, \nabla^2 \mathbf{Q}^{**}, ...) \label{eq:PDE-source}
\end{align}    
\end{subequations}

In the above splitting scheme, $\mathbf{Q}^n, \mathbf{Q}^*, \mathbf{Q}^{**} \text{ and } \mathbf{Q}^{n+1}$ denote the old, the two intermediate and the new solution for each time step $n$. The source term is divided into a pressure and friction part $\mathbf{S} ^P$ and $\mathbf{S}^F$, where all terms involving pressure $\mathbf{P}$ are absorbed in $\mathbf{S}^P$.

Note that the pressure $\mathbf{P}$ \eqref{eq:PDE-non-hydrostatic} is obtained by first inserting $\mathbf{Q}^{**}$ from \eqref{eq:PDE-non-hydrostatic-update} into \eqref{eq:PDE-non-hydrostatic} to obtain a Poisson-type equation depending only on known values of $\mathbf{Q}^*$ and unknowns $\mathbf{P}$. After solving for $\mathbf{P}$, $\mathbf{Q}^{**}$ is obtained by the update in \eqref{eq:PDE-non-hydrostatic-update}.

Based on \eqref{eq:PDE-hyperbolic} to \eqref{eq:PDE-source}, we identify the necessary code building blocks

\begin{itemize}
    \item \code{step\_hyperbolic} for \eqref{eq:PDE-hyperbolic},
    \item \code{newton\_solver} for \eqref{eq:PDE-non-hydrostatic},
    \item \code{step\_source} for \eqref{eq:PDE-non-hydrostatic-update} and \eqref{eq:PDE-source},
    \item \code{step\_source\_implicit} for \eqref{eq:PDE-source} in implicit mode,
    \item \code{boundary\_operator},
    \item and \code{compute\_derivative} for the construction of $\nabla^k \mathbf{Q}$ and $\nabla^l \mathbf{P}$,
\end{itemize}
for the implementation of the predictor-corrector scheme.

In the current implementation, we provide building blocks that target numerical schemes on collocated grids. Although the \code{mesh} class provides access to nodes and cell faces, future work is required to add convenient functions, such as projection schemes between cells, nodes, and faces, to facilitate the simple construction of methods on staggered grids.

The above building blocks can be used to construct many different solvers beyond the transient semi-implicit scheme described in \eqref{eq:PDE-hyperbolic}-\eqref{eq:PDE-source}. Fully implicit schemes are similarly possible, mainly due to the flexibility of the residual definition of the Newton solver. 
In the current release, we provide the solvers shown in Figure \ref{fig:building_blocks}.

\subsection{Comparable, accessible and simple}

The numerical models for advanced depth-averaged free-surface models are typically challenging to derive, modify, and interpret. In our opinion, the main complication stems from the fact that the five main flow variables, $h, (u,\, v,\, w)$ and $p$ for the water height, the velocity, and the pressure are described with different sets of variables, e.g., $h, (u0, \,u1, \,v0, \,v1, \,w0, \,w1, \,w2), (p0, \,p1, \,p2)$ for the VAM model \cite{Escalante_Morales_De_Luna_Cantero-Chinchilla_Castro-Orgaz_2024}. Without expertise and a careful examination of the underlying derivation, the reader can quickly lose sight of the connection between variables and physical intuition.

To promote the use of depth-averaged free-surface flow models, one core design principle of our software is to ensure comparability between different models, allowing for the accessibility of the software for any user with minimal effort, and simplifying the initial experience with new models to a level where no prior knowledge of the PDE is required.
% \todo[inline]{maybe add the sentence that briefly lists the specific steps? like "To this end, we provide a convenient projection back to the physical flow variables... }

\subsubsection{Postprocessing} \label{sec:postprocessing}

Use cases such as model selection for an engineering application or benchmarking newly developed models benefit from existing reference implementations and require their comparability. To achieve this, a dedicated modeling software for free-surface flow models is useful. 

For this reason, each model contains a function \code{interpolate\_3d} that can be implemented to recover the flow variables $h, (u, v, w)$ and $p$ of the three-dimensional incompressible Navier-Stokes or RANS setting. The function can be augmented with auxiliary variables of interest. 

This simple post-processing feature facilitates connecting the large community of scientists working with three-dimensional flow solvers with depth-averaged free-surface models. 

We store the solutions during the solver run in a custom \textit{hdf5}-format, to allow restarts of the simulation and, in principle, the compatibility between multiple numerical backends and programming languages such as \textit{Python} and \textit{C++}. For further post-processing, we currently support conversion to the \textit{VTK} format, which can, for example, be opened in \textit{ParaView}, a popular open source scientific visualization software \cite{paraview}.

\subsubsection{Cross-platform Open-Source Software}

Zoomy is published as open source software available at \url{https://github.com/mbd-rwth/Zoomy}.
We make our project accessible by supporting installations via the \textit{conda} package manager \cite{anaconda_software_2016} or manual installation from source. Additionally, we support prebuilt container environments for Apptainer \cite{apptainer} and Docker \cite{merkel2014docker}.

\subsubsection{Cross-platform Graphical User Interface}
\begin{figure}
    \centering
    \includegraphics[width=1\linewidth]{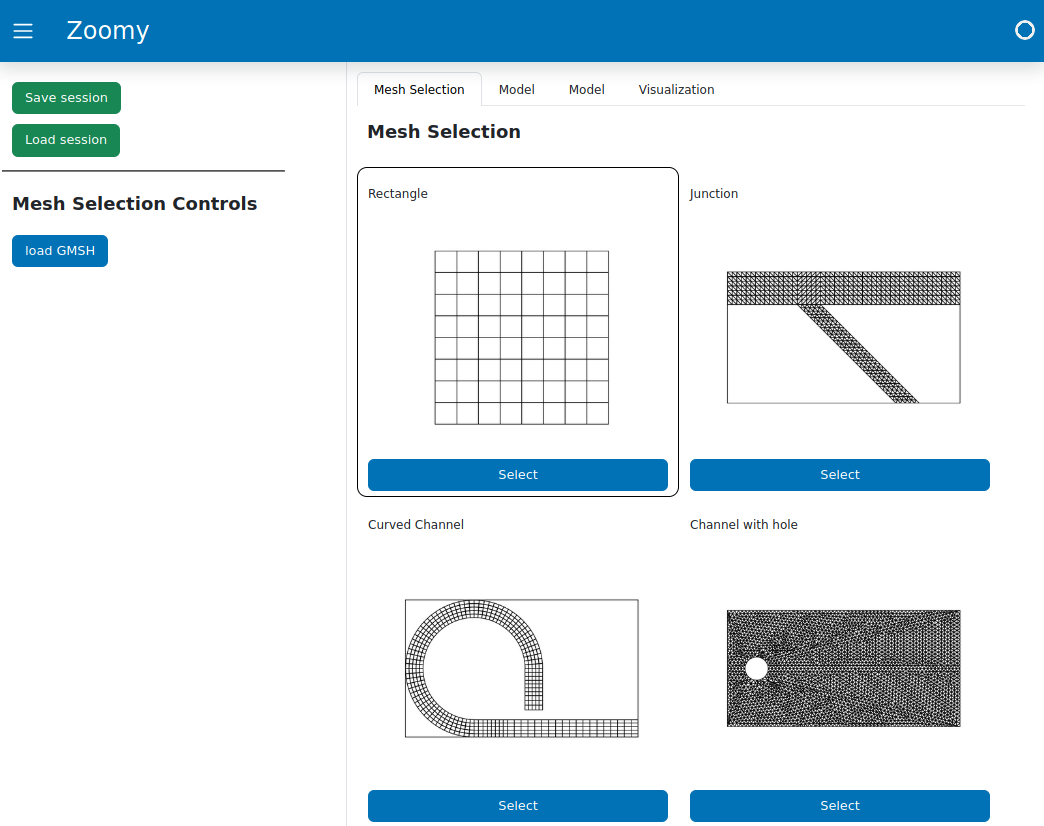}
    \caption{Graphical User Interface (GUI). Displays the card system for the mesh selection. The same card system is used to select models, numerical solvers and visualization tools. Customizations of each selected model are possible using advanced settings or by directly editing the code of each class in the GUI's code editor.}
    \label{fig:gui}
\end{figure}

Providing tutorials in \textit{Jupyter Notebooks}, e.g. hosted on cloud services such as Binder \cite{binder} or Google Collab \cite{googlecollab} allows users a simple access to new software with minimal initial investment of time.
Building on the same idea, we aim to create a Graphical User Interface (GUI) that serves as a simple entry point for users to explore our software as well as depth-averaged free-surface models in general. In contrast to \textit{Jupyter Notebooks}, our goal is for the first interaction between the user and our software to be as simple as possible, which requires a math-free and code-free experience.
On the other hand, the GUI should be more than a demonstrator but allow the exploration and modification of existing models. 

This led to the development of a simple GUI using the Python package \code{panel}. \code{Panel} is a tool for developing dashboards and other graphical user interface applications. Compared to other solutions, we chose \code{panel} as our GUI platform because it offers the following features:

\begin{itemize}
    \item cross-platform and cloud-service capable, as \code{panel} can run using a \textit{client-server} architecture
    \item allows native support of common plotting libraries such as Matplotlib
    \item allows to integrate more complex visualization software, such as ParaView Lite \cite{paraview} 
    \item ships a ready-to-use code editor interface
\end{itemize}

Note that the last point is crucial from our perspective, primarily because it eliminates the need for us to build and maintain a large number of input fields, such as sliders or text fields. Instead, it allows us to expose the user to two different \textit{views}, the minimal and simple \textit{card view} and the fully featured \textit{code view}.

An example of the \textit{card view} is presented in Figure \ref{fig:gui}. The \textit{card view} displays multiple \textit{cards} that allow the user to select between different existing meshes. 
The workflow for configuring and running a simulation is organized into multiple \textit{tabs}: the \textit{mesh page}, \textit{model page}, \textit{solver page} and \textit{visualization page}. Each \textit{page} holds its own set of \textit{cards} to select from. 
The sidebar on the left of the \textit{card view} can be used to place classical GUI components such as buttons, sliders, or text fields. In this initial version of the GUI, we have implemented only a minimal set of additional elements, as the simulation remains fully configurable in the \textit{code view}.

The \textit{code view} is accessible by clicking on the respective button on the card. This opens a new \textit{page} with the code editor showing the respective code in the underlying Python file. 

\begin{figure}
    \centering
    \includegraphics[width=1\linewidth]{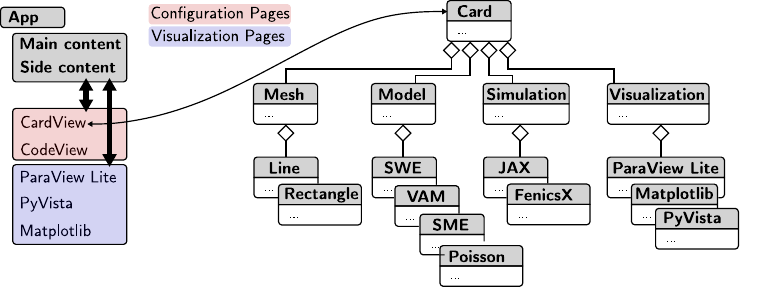}
    \caption{GUI Design.}
    \label{fig:GUI-design}
\end{figure}

Using the components provided by \code{panel}, we have developed the \textit{pages} and \textit{card view}, in particular by introducing the classes \code{Card}, \code{CardManager}, \code{Page} and \code{PageManager}. The interaction between these classes is summarized in the design of our GUI, as shown in Fig. \ref{fig:GUI-design}. 

We currently support three different visualization backends: \textit{ParaView Lite} and \textit{PyVista} \cite{sullivan2019pyvista} for 2D and 3D data and \code{Matplotlib} \cite{matplotlib} for 1D data.

The GUI also serves as a configurator that produces the complete configuration of the test case as a Python script. The cases are therefore reproducible and automatically compatible with the command-line tool.

\section{Modeling Examples}\label{sec:examples}

To demonstrate the various use cases and components established in the preceding sections, we will showcase different models that can be generated using our software. We will begin with the classical SWE in one and two dimensions and demonstrate the transient hyperbolic solver. We will follow up with a 1D test case solving the Poisson equation to show how the Newton solver can be used.  Next, we will combine the two solvers to build a predictor-corrector scheme for \eqref{eq:PDE-hyperbolic}-\eqref{eq:PDE-source} to solve for the non-hydrostatic VAM model. Lastly, we will show how the model hierarchy for the hydrostatic SME with a variable polynomial degree can be implemented.

\subsection{Shallow Water Equations in 1D and 2D}

In this test case, we demonstrate
\begin{itemize}
    \item how the symbolic model definition works,
    \item the construction of multi-dimensional models,
    \item the eigenvalues can be symbolically computed,
    \item the bridge between the symbolic and numerical model,
    \item the correctness of the hyperbolic solver and
    \item how GMSH meshes can be used.
\end{itemize}
This test case can be found in the folder \code{tutorials/swe/simple.ipynb}.

Following our general PDE template \eqref{eq:PDE-structure}, the SWE without bottom topography and friction read

\begin{align}
    \partial_t \mathbf{Q} + \nabla \cdot \underline{\underline{F}}(\mathbf{Q}) = \mathbf{0}
\end{align}

with

\begin{align}
    \underline{\underline{F}}(\mathbf{Q})= \left[ \mathbf{F}_x(\mathbf{Q}), \mathbf{F}_y(\mathbf{Q}) \right] = 
    \left[ 
    \begin{pmatrix}
        hu \\ hu^2 + g h \\ huv
    \end{pmatrix}, 
    \begin{pmatrix}
        hv \\ huv \\ hv^2 + gh
    \end{pmatrix}
    \right] ,
\end{align}

where $h, u, v$ denote the height of the water and the velocity x and y, respectively. $g$ denotes the gravitational acceleration in the z-direction.

\begin{minipage}[h]{1.\linewidth}
\begin{python}
class ShallowWater(Model):
    """
    A 1D and 2D implementation of the Shallow Water Equations
    """

    def flux(self):
        # construct a (dimension, number of fields) matrix
        flux = Matrix([[0 for i in range(self.n_fields)] for j in range(self.dimension])
        h = self.variables[0]
        #hU = [hu] in 1D and [hu, hv] in 2D
        hU = self.variables[1:1+self.dimension]
        I = Identity(self.dimension)
        param = self.parameters
        flux[:, 0] = hU
        flux[:, 1:1+self.dimension] = hU * hU.T + 1/2*g*h**2 * I
        # the output expects a list of (n_fields, 1) matrices, 
        # e.g. [F_x] in 1D and [F_x, F_y] in 2D
        return [flux[i] for i in range(self.dimension)]
\end{python}
\end{minipage}

By omitting the constructor, initial and boundary code for clarity, the implementation of the mathematical model is already complete. 
Under the hood, the \code{nonconservative\_matrix}, \code{source} and \code{residual} functions are initialized with zero. Based on the \code{flux} ($\underline{\underline{F}}$) and \code{nonconservative\_matrix} ($\underline{\underline{\underline{NC}}}$), the eigenvalues can automatically be computed by solving the generalized eigenvalue problem.

\begin{align}
    \left| \frac{\partial \underline{\underline{F}}}{\partial \mathbf{Q}} \cdot \mathbf{n} + \underline{\underline{\underline{NC}}} \cdot \mathbf{n} + \lambda \underline{\underline{I}} \right| \stackrel{!}{=} 0
\end{align}

where $\mathbf{n}$ is a symbolic directional vector.

Note that the eigenvalues calculated by \code{SymPy} are not necessarily in a form that is advantageous for the numerical discretization. In our case, the eigenvalues without any additional simplifications or enforcing assumptions read

\begin{align}
\begin{aligned}
    \lambda_0 & = \frac{h\mathbf{u} \cdot \mathbf{n}}{h} \\
    \lambda_{1, 2} &=  \frac{h \,h\mathbf{u} \cdot \mathbf{n} \pm \sqrt{g h^5} \sqrt{\mathbf{n}^2}}{h^2} \quad .
\end{aligned}
\end{align}

This is mathematically correct, but additional processing to simplify $\mathbf{n}^2 = 1$ and reduce the power $h^5$ in the square root is advantageous. In future releases, we will implement additional automatic simplifications. However, manual simplifications are already possible by overloading the existing functionality of the \code{eigenvalues} function before the code is transformed into the \textit{numerical layer}.

As mentioned in Section \ref{sec:postprocessing}, we encourage the implementation of the \code{interpolate\_3d} function to make the results of the model easily comparable with other free-surface flow models.
For the SWE, the underlying assumptions of a constant velocity profile in the z direction, as well as hydrostatic pressure, yield the following lifting.

\begin{align}
    \begin{aligned}
        \rho(t,x,y,z) &= \rho_{water} * ((h(t,x,y)^{SWE}-z) \geq 0 \, ? \, 1 \, : \, 0) \\
        u(t,x,y,z) &= u^{SWE}(t,x,y) \\
        v(t,x,y,z) &= v^{SWE}(t,x,y) \\
        w(t,x,y,z) &= - h  \partial_x u^{SWE}(t,x,y) -h  \partial_y v^{SWE}(t,x,y) \\
        p(t,x,y,z) &= \rho_{water} \, g  \,(\max(h(t,x,y)^{SWE} - z, 0))
    \end{aligned}
\end{align}

Note that the above formulation changes if a bottom topography or air phase is explicitly considered.

The implementation closely follows the symbolic definition

\begin{python}
    def interpolate_3d(self):
        param = self.parameters
        h = self.variables[0]
        u = self.variables[1]/h
        dudx = self.aux_variables.dudx
        v = 0
        dvdy = 0
        if self.dimension == 2:
            v = self.variables[2]
            dvdy = self.aux_variables.dvdy
        rho = param.rho * Piecewise((1, h - z), (0, True))
        w = - h * dudx - h * dvdy
        p = param.rho * param.g * Piecewise((h-z, h-z > 0), (0, True)) 
        return rho, u, v, w, p
\end{python}

with the main difference that the spatial derivatives are represented as symbolic auxiliary variables. 

As a numerical solver, we can inherit the implementation of the \code{TransientHyperbolic} solver. The only customization required is to overload the \code{update\_qaux} function to compute the gradients of the velocity field:

\begin{python}
    class SWESolver(TransientHyperbolic):
    """
    A Shallow Water Equations solver for 1D and 2D problems
    """
        def update_qaux(self, Q, Qaux, Qold, Qauxold, mesh, 
                        model, parameters, time, dt):
            h = Q[0]
            u = Q[1]/h
            dudx  = compute_derivatives(
                    u, mesh, 
                    derivatives_multi_index=([[1, 0]])
                    )[:,0]
            # this is jax syntax for Qaux[0] = dudx
            Qaux = Qaux.at[0].set(dudx)
            if model.dimension == 2:
                v = Q[2]/h
                dvdy  = compute_derivatives(
                        v, mesh,
                        derivatives_multi_index=([[0,1]])
                        )[:,0]
                # this is jax syntax for Qaux[1] = dvdy
                Qaux = Qaux.at[1].set(dvdy)
            return Qaux
\end{python}

With all components in place, a runner for the final 2D simulation now reads

\begin{python}
    ...
    model = ShallowWater(
        dimension=2,
        aux_fields=['dudx', 'dvdy'],
        parameters={'g': 9.81, 'rho': 1000},
        boundary_conditions=bcs,
        initial_conditions=ic,
    )

    main_dir = os.getenv("Zoomy")
    mesh = petscMesh.Mesh.from_gmsh(
        os.path.join(main_dir, "meshes/quad_2d/mesh_coarse.msh")
    )
    solver = SWESolver(
            compute_dt=timestepping.adaptive(CFL=0.45),
            time_end=0.1,
            )
    Qnew, Qaux = solver.solve(mesh, model, settings)

    # generates the VTK solution of the SWE fields
    io.generate_vtk(settings.output_dir)
    # generates the VTK solution of the using interpolate_3d
    io.generate_lifted_vtk(settings.output_dir)
\end{python}

\begin{figure}
    \centering
    \includegraphics[width=1.\linewidth]{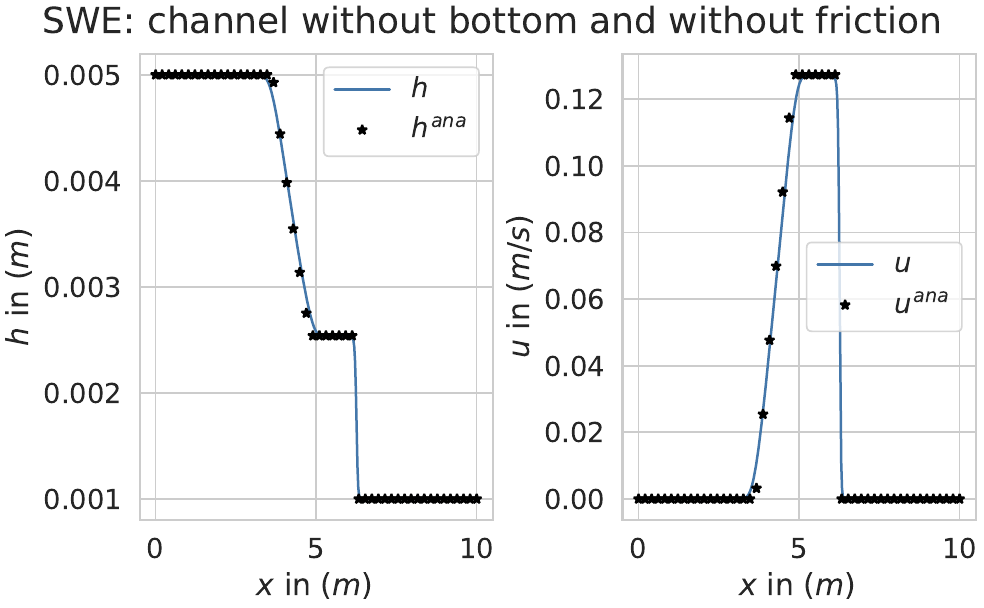}
    \caption{Comparison of our numerical solution with the analytical solution. Our two-dimensional solution is sliced along $y=1$ for the comparison.}
    \label{fig:result-swe}
\end{figure}

Finally, the numerical solution and comparison to the test case \textit{Dam break on a wet domain without friction} from \cite{Delestre_Lucas_Ksinant_Darboux_Laguerre_Vo_James_Cordier_2013} is shown in Figure \ref{fig:result-swe}.

\subsection{Steady Poisson Equation in 1D}

As the solution of non-hydrostatic flow equations requires the solution to a Poisson-type system, we introduce this building block using the simple steady Poisson equation in 1D.

In this test case, we demonstrate
\begin{itemize}
    \item how the residual definition is used to construct an elliptic solver
    \item the correctness of the \code{SteadyResidual} solver
\end{itemize}
This test case can be found in the folder \code{tutorials/poisson/simple\_1d.ipynb}.

Only considering (\ref{eq:PDE-structure}.2) of our general PDE template, the Poisson equation 

\begin{align}\label{eq:poisson}
\begin{cases}
    \partial_{xx} T(x) = 2 \text{ for } x \in \Omega = [0, 1] \\
    T(0) = 1\\ 
    T(1) = 2
\end{cases}  
\end{align}

on the domain $x=[0, 1]$ can be brought into residual form and used to define the model 

\begin{python}
class Poisson(Model):
    def residual(self):
        R = Matrix([0 for i in range(self.n_fields)])
        T = self.variables[0]
        ddTdxx = self.aux_variables.ddTdxx
        param = self.parameters

        R[0] = - ddTdxx + 2
        return R
\end{python}

together with the Dirichlet boundary conditions

\begin{python}
bcs = BC.BoundaryConditions( [
    BC.Lambda(physical_tag='left', 
        prescribe_fields={0: lambda t, x, dx, q, qaux, p, n: 1.}
        ),
    BC.Lambda(physical_tag='right', 
        prescribe_fields={0: lambda t, x, dx, q, qaux, p, n: 2.}
        ),
])
\end{python}

The derivative $\partial_{xx} T$ are defined in the update of the auxiliary variables:

\begin{minipage}[h]{1.\linewidth}
\begin{python}
class PoissonSolver(SteadyResidual):
    def update_qaux(self, Q, Qaux, Qold, Qauxold, 
                    mesh, model, parameters, time, dt
                    ):
        T = Q[0]
        ddTdxx  = compute_derivatives(T, mesh, 
                    derivatives_multi_index=([[2]])
                    )[:,0]
        # jax syntax for Qaux[0] = ddTdxx
        Qaux = Qaux.at[0].set(ddTdxx)
        return Qaux
\end{python}
\end{minipage}

\begin{minipage}[htb]{0.5\linewidth}
\begin{python}
def default_residual(Q):
    Qaux = self.update_qaux(Q, Qaux, Qold, Qauxold, 
                            mesh, model, parameters, time, dt
                            )
    Q = boundary_operator(time, Q, Qaux, parameters)
    res = model.residual(Q, Qaux, parameters)
    # jax syntax for res[:, mesh.n_inner_cells:] = 0.
    res = res.at[:, mesh.n_inner_cells:].set(0.)
    return res
\end{python}
\end{minipage}

This is all that needs to be done, except for writing a runner script. In particular, no additional information regarding the assembly of the linear system is required, as a matrix-free GMRES linear solver is used internally. 
Construction of the residual Jacobian needs no user specification, as the Jacobian is built using automatic differentiation. 

The GMRES solver takes a \code{default\_residual} implementation if the user does not specify a \code{custom\_residual} function.

\begin{figure}
    \centering
    \includegraphics[width=0.75\linewidth]{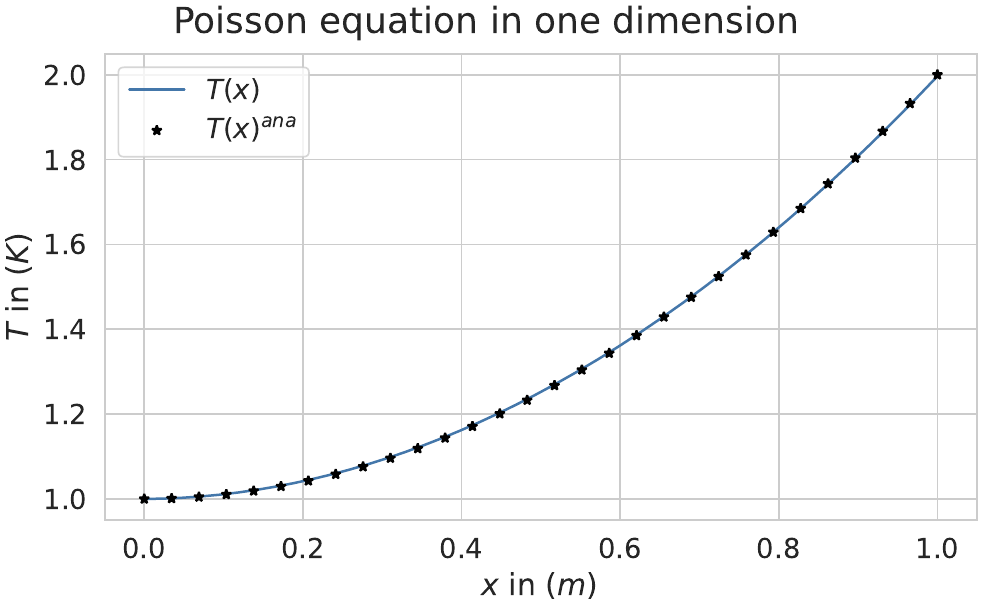}
    \caption{Comparison between the numerical and analytical result of the Poisson problem.}
    \label{fig:poisson}
\end{figure}

To avoid considering ghost cells during the construction of the Jacobian using automatic differentiation, we set the residual at the ghost cells to zero (\textit{line 8}). Figure \ref{fig:poisson} shows the comparison between the simulation and the analytical solution $T(x) = x^2 + 1$ of \eqref{eq:poisson}.

\subsection{Unsteady Vertically Averaged Momentum Equations in 1D}\label{sec:VAM}

In this test case, we demonstrate
\begin{itemize}
    \item how \eqref{eq:PDE-hyperbolic}-\eqref{eq:PDE-source} can be implemented with a multi-model and multi-solver approach to implement the predictor-corrector scheme,
    \item the correctness of the numerical solver in comparison with the analytical solution 
\end{itemize}
This test case can be found in the folder \code{tutorials/vam/simple\_1d.ipynb}.

The key idea behind the derivation of the VAM model is the depth averaging of the RANS equations. Different derivations of the VAM model exist \cite{Khan_Steffler_1996a, Escalante_Morales_De_Luna_Cantero-Chinchilla_Castro-Orgaz_2024}. In this paper, we are concerned with the version recently derived in \cite{Escalante_Morales_De_Luna_Cantero-Chinchilla_Castro-Orgaz_2024}, as it uses a nearly identical approach to the one taken in the derivation of the SME in \cite{Kowalski_Torrilhon_2019} demonstrated in the next example. The primary difference between the two models is that the VAM model incorporates non-hydrostatic contributions and is numerically more complex. The SME has been used to explore the model hierarchy itself, which poses challenges in its implementation and analysis.

The 1D VAM model in \cite{Escalante_Morales_De_Luna_Cantero-Chinchilla_Castro-Orgaz_2024} assumes that
\begin{itemize}
    \item $u$ is linear,
    \item $w$ is quadratic,
    \item $p$ is quadratic
\end{itemize}
and the functions are constructed using Legendre polynomials as a basis.

The VAM equations, without considering friction, can be written in the form \eqref{eq:PDE-hyperbolic}-\eqref{eq:PDE-source} by identifying

\begin{align}
    \begin{aligned}
     \begin{aligned}
         \mathbf{Q} =& 
        \begin{pmatrix}
            h \\ h u_0 \\ hu_1 \\ hw_0 \\ hw_1
        \end{pmatrix} \\
        \underline{\underline{NC}} =&
        \begin{pmatrix}
            0 & 0 & 0 & 0 & 0 \\
            g \, h & 0 & 0 & 0 & 0 \\
            0 & 0 & -u_0 & 0 & 0 \\
            0 & 0 & 0 & 0 & 0 \\
            0 & 0 & \frac{1}{5} w_2 - w_0 & 0 & 0
        \end{pmatrix}
     \end{aligned} \quad
     \begin{aligned}
         \mathbf{F} = &
        \begin{pmatrix}
            hu_0 \\ hu_0^2 + \frac{1}{3} hu_1^2 \\ 2 hu_0 u_1 \\ h u_0 w_0 + \frac{1}{3} hu_1 w_1 \\ hu_0 w_1 + u_1(hw_0 + \frac{2}{5} h w_2)
        \end{pmatrix} \\
        \mathbf{S}^P = &
        \begin{pmatrix}
            0 \\
            \partial_x(hp_0) + 2 p_1 \partial_x b \\
            -(3 p_0 - p_1) \partial_x h  - 6(p_0 - p_1)\partial_x b\\
            2 p_1 \\
            6(p_0 - p_1)
        \end{pmatrix} 
        \end{aligned} \\
    \end{aligned}
\end{align}\label{eq:vamQ}

\begin{align}
    \begin{aligned}
        R_0 &= h \partial_x u_0 + \frac{1}{3} \partial_x (h u_1) + \frac{1}{3} u_1 \partial_x h + 2 (w_0 - u_0 \partial_x b) \\
        R_1 &= h \partial_x u_0 + u_1 \partial_x h +2 (u_1 \partial_x b-w_1)
    \end{aligned}
\end{align}\label{eq:vamP}

\begin{align}
    w_2 = -(w_0 + w_1) + (u_0 + u_1) \partial_x b \quad .
\end{align}\label{eq:vamW2}

The definition of the first model \code{VAMPredictor} can be done similarly to the above examples, once we identify
$\mathbf{Q}_{aux} = \left(w_2,\, p_0, \,p_1, \partial_x(h\, p_0), \partial_x h, \, \partial_x b \right)^T$.

The definition of the second model \code{VAMCorrector} requires us to analytically substitute \eqref{eq:PDE-non-hydrostatic-update} in \eqref{eq:PDE-non-hydrostatic} to generate a Poisson-type equation.

This can be done symbolically with a code similar to

\begin{python}
S = Matrix([[0, 
           (h * p0).diff(x) + 2*p1*b.diff(x),
           (h * p1).diff(x) -(3*p0 - p1)*h.diff(x) - 6*(p0-p1) * b.diff(x),
            - 2*p1,
           6 * (p0-p1)
          ]]).T
Uk = Matrix([[ h, h * oldu0, h*oldu1, h*oldw0, h*oldw1]]).T
U = Uk - dt * S
h = U[0]
u0 = U[1]/h
u1 = U[2]/h
w0 = U[3]/h
w1 = U[4]/h
R1 = h*u0.diff(x) + 1/3 * (h*u1).diff(x) + 1/3 * u1 * h.diff(x) + 2*(w0 - u0 * b.diff(x) + h.diff(t).diff(x).diff(y))
R1 = replace_variables(R1, U)
R2 = h * u0.diff(x) + u1*h.diff(x) + 2*(u1*b.diff(x) - w1)
R2 = replace_variables(R2, U)
\end{python}
where \code{replace\_variables} substitutes \code{U} in the expressions \code{R1} and \code{R2}, resulting in the Poission type equation depending on unknowns $(\partial_{xx} p_0, \partial_{xx} p_1, \partial_x p_0, \partial_x p_1, p0, \,p1)$ and coefficients of known variables.

For the numerical solution of the model, we now need to construct two solvers, \code{VAMPredictorSolver} based on the \code{HyperbolicSolver} solver and \code{VamCorrectorSolver} based on the \code{SteadyResidualSolver} solver.
In both cases, only the appropriate \code{update\_qaux} functions need to be implemented.

The final solver is now a composite of the two solvers. This solver can be inherited from the base class \code{Solver}, as we need to implement our own time loop. In this time loop, we perform the following: 

\begin{figure}[!htb]
    \centering
    \includegraphics[width=1\linewidth]{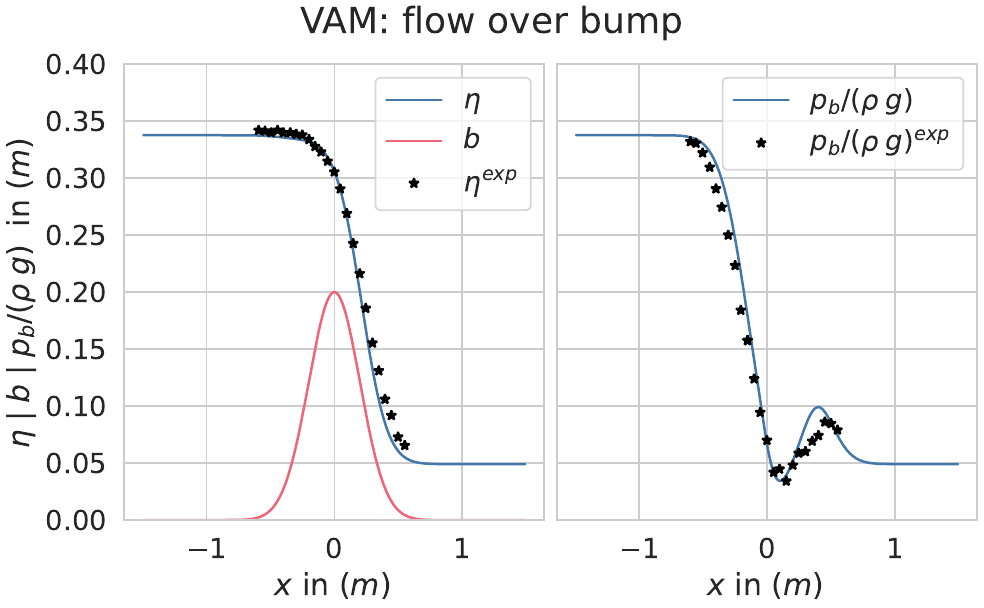}
    \caption{Comparison of the VAM model and the analytical solution for the test case described in \cite{Sivakumaran_Tingsanchali_Hosking_1983}.}
    \label{fig:vam}
\end{figure}

\begin{python}
    ...
    Q = boundary_operator_Q(time, Q, Qaux, parameters)
    Qaux = predictor.update_qaux(
        Q, Qaux, Qold, Qauxold, mesh, model, parameters time, dt
        )
    step_hyperbolic(time, Q, Qaux, parameters, dQ)
    # Copy Q^* into Paux to make it available in the corrector solver
    Paux = Paux.at[0:6].set(Q[0:6])
    P = corrector.solve(P)
    # Copy P to Qaux
    Qaux = Qaux.at[0:2].set(P)
    Q = boundary_operator_Q(time, Q, Qaux, parameters)
    Qaux = predictor.update_qaux(
        Q, Qaux, Qold, Qauxold, mesh, model, parameters time, dt
    )
    Q = step_source(mesh, model)
\end{python}

The main addition is the data transfer from $\mathbf{Q} \rightarrow \mathbf{P}_{aux}$ and back from $\mathbf{P} \rightarrow \mathbf{Q}_{aux}$.

Figure \ref{fig:vam} shows the numerical solution and experimental solution for a test case specified in \cite{Sivakumaran_Tingsanchali_Hosking_1983}.

% Lastly, we can perform a linear stability analysis almost automatically due to \code{SymPy} and our implementation of the convenience function \code{linear\_stability\_analysis}. The function requires an expansion of our state variables $\mathbf{Q}$ in terms of the smallness parameter $\epsilon$ and substitution rules for the variables $\mathbf{Q}_{aux}$ as a dictionary.

% Linearizing around a lake-at-rest solution yields

% \todo[inline]{TODO}

\subsection{Shallow Moment Equations in 2D with friction}

In this test case, we demonstrate
\begin{itemize}
    \item the generation and numerical solution of the model hierarchy
    \item how the material closure can be inserted
    \item how the dimensional reconstruction.  
\end{itemize}
This test case can be found in the folder \code{tutorials/sme/simple\_2d.ipynb}.
In contrast to the VAM model in Section \ref{sec:VAM}, the 2D SME model \cite{Steldermann_Torrilhon_Kowalski_2023} assumes that
\begin{itemize}
    \item $u(t,x,y,\zeta)=\sum{i=0}^N \alpha_i(t,x,y) \phi_i(\zeta)$ where $\phi_i$ are Legendre polynomials up to degree $N$,
    \item $w$ is implicitly defined by the mass balance and
    \item $p$ is hydrostatic.
\end{itemize}
The PDE system is given by

\begin{align*}
    & \mathbf{Q} = 
    \begin{pNiceArray}{l}[last-col]
        h & \textcolor{blue}{\downarrow 1}\\
        h \alpha_k & \textcolor{blue}{\downarrow N}\\
        h \beta_k & \textcolor{blue}{\downarrow N}
    \end{pNiceArray}  \\[0.2cm]
    & \underline{\underline{F}}(\mathbf{Q}) = 
    \begin{pmatrix}
    \begin{pNiceArray}{l}[last-col]
        h \alpha_0 & \textcolor{blue}{\downarrow 1}\\
        \sum_{i,j=0}^{N} h \alpha_i \alpha_j  A_{ijk} / M_{kk} +  \delta_{k0} \frac{g}{2} h^2 & \textcolor{blue}{\downarrow N}\\
        \sum_{i,j=0}^{N} h \alpha_i \beta_j  A_{ijk} / M_{kk} & \textcolor{blue}{\downarrow N}\\
    \end{pNiceArray},
    \begin{pNiceArray}{l}[last-col]
        h \beta_0 & \textcolor{blue}{\downarrow 1} \\
        \sum_{i,j=0}^{N} h \beta_i \alpha_j A_{ijk} / M_{kk} & \textcolor{blue}{\downarrow N} \\
        \sum_{i,j=0}^{N} h \beta_i \beta_j  A_{ijk} / M_{kk} +  \delta_{k0} \frac{g}{2}h^2 & \textcolor{blue}{\downarrow N} \\
    \end{pNiceArray}
    \end{pmatrix} \\[0.2cm]
    & \underline{\underline{\underline{B}}}(\mathbf{Q})=
    \left( \rule{0pt}{2.2em} \right.
    \begin{pNiceArray}{lll}[first-row, last-col]
      \textcolor{blue}{\rightarrow 1} & \textcolor{blue}{\rightarrow N} & \textcolor{blue}{\rightarrow N} & \\
      0 &  0  & 0 & \textcolor{blue}{\downarrow 1} \\
      0 & \left( u_m \delta_{kl}  - \alpha_j B_{jlk} / M_{lk}  \right) & 0 & \textcolor{blue}{\downarrow N} \\
        0 & \left( v_m \delta_{kl}  - \beta_j B_{jlk} / M_{lk} \right) & 0 & \textcolor{blue}{\downarrow N}
    \end{pNiceArray}, 
    \begin{pNiceArray}{lll}[first-row, last-col]
      \textcolor{blue}{\rightarrow 1} & \textcolor{blue}{\rightarrow N} & \textcolor{blue}{\rightarrow N} & \\
      0 &  0  & 0 & \textcolor{blue}{\downarrow 1} \\
      0 &  0 & \left( u_m \delta_{kl}  - \alpha_j B_{jlk} / M_{lk} \right) & \textcolor{blue}{\downarrow N} \\
        0 & 0 & \left(  v_m \delta_{kl}  - \beta_j B_{jlk} / M_{lk} \right) & \textcolor{blue}{\downarrow N}
    \end{pNiceArray}
    \left. \rule{0pt}{2.2em} \right)
    \\[0.2cm]
    & \mathbf{S}(\mathbf{Q})= 
    \begin{pNiceArray}{l}[last-col]
        0 & \textcolor{blue}{\downarrow 1} \\
        - \left.\left( \rho^{-1} \tilde{\sigma}_{xz} \, \phi_k \right) \right\rvert_{\zeta=0} / M_{kk} + \left\langle \rho^{-1} \tilde{\sigma}_{xz} , \phi_k' \right\rangle / M_{kk} & \textcolor{blue}{\downarrow N}\\
        - \left.\left( \rho^{-1} \tilde{\sigma}_{yz} \, \phi_k \right) \right\rvert_{\zeta=0} / M_{kk}+ \left\langle \rho^{-1} \tilde{\sigma}_{yz} , \phi_k' \right\rangle / M_{kk} & \textcolor{blue}{\downarrow N}\\
    \end{pNiceArray}
\end{align*}
with $k,l=0,\ldots, N$ generate the moment system. The blue arrows indicate the number of equations in each block. 

To be concise, we focus only on the implementation of the source term $\mathbf{S}(\mathbf{Q})$, which has not yet been addressed in any other example. The source term consists of a boundary closure term for the bottom friction and a bulk friction term and does not yet assume any material model.

To demonstrate a simple set of closure relations, we assume a slip boundary condition and a Newtonian fluid for the bottom friction and bulk friction terms, respectively. The closure then reads

\begin{align}
    \begin{aligned}
        & \left. \tilde{\sigma}_{xz} \right\rvert_{\zeta=0} = C \left. u \right\rvert_{\zeta=0} \\
        & \left. \tilde{\sigma}_{yz} \right\rvert_{\zeta=0} = C \left. v \right\rvert_{\zeta=0} \\
        & \tilde{\sigma}_{xz} = \nu \left( \partial_z u \right) \\
        & \tilde{\sigma}_{yz} = \nu \left( \partial_z v \right) \quad ,
    \end{aligned}
\end{align}

with $C$ the scaled slip-length and $\nu$ the dynamic viscosity. Note that in the bulk closure, the terms $\partial_x w$ and $\partial_y w$ were based on a scaling argument in \cite{Kowalski_Torrilhon_2019}.

The bottom friction terms can be easily implemented, as the velocity at the bottom is readily available:

\begin{python}
    def slip_boundary_condition(self):
        out = Matrix([0 for i in range(self.n_fields)])
        offset = self.levels+1
        h = self.variables[0]
        ha = self.variables[1 : 1 + self.levels + 1]
        hb = self.variables[1+offset : 1+offset + self.levels + 1]
        p = self.parameters
        ub = 0
        vb = 0
        phi_0 = [self.basismatrices.eval(i, 0.0) for i in range(self.levels + 1)]
        for i in range(1 + self.levels):
            ub += ha[i]*phi_0[i] / h
            vb += hb[i]*phi_0[i] / h
        for k in range(1, 1 + self.levels):
            out[1 + k] += (
                -1.0 * p.C /  p.rho * ub * phi_0[k] / self.basismatrices.M[k, k]
            )
            out[1+offset+k] += (
                -1.0 * p.C / p.rho * vb *phi_0[k] / self.basismatrices.M[k, k]
            )
        return out
\end{python}

\begin{figure}[!htb]
    \centering
    \includegraphics[width=1\linewidth]{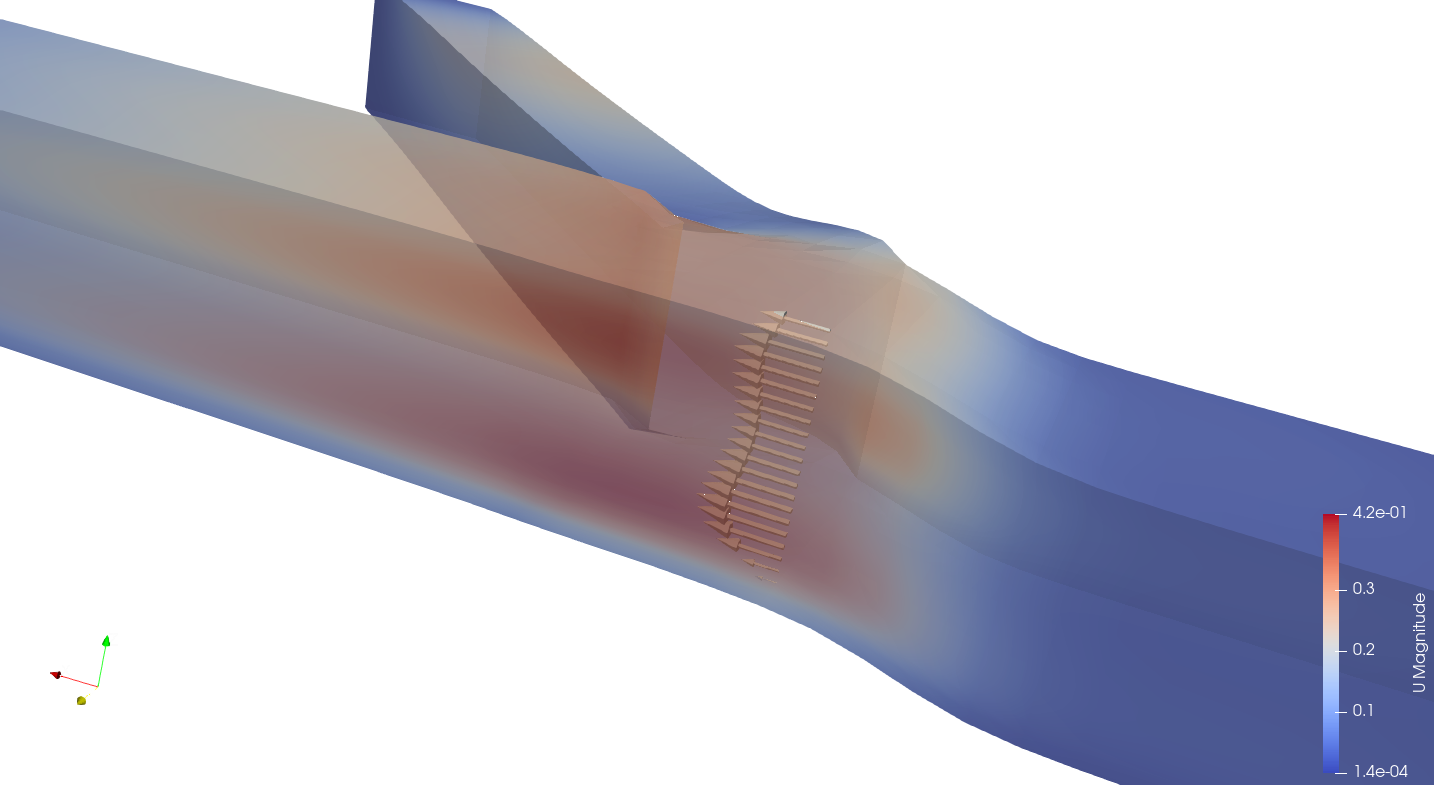}
    \caption{3D reconstruction of a two-dimensional simulation using the SME with polynomials of order 4.}
    \label{fig:sme}
\end{figure}

The bulk friction term requires symbolic integration. For a Newtonian fluid, the term can be simplified 

\begin{align}
    \begin{aligned}
       &  \left\langle \rho^{-1} \tilde{\sigma}_{xz} , \phi_k' \right\rangle / M_{kk} = 
       \frac{\nu}{\rho} \sum_{i=0}^N \alpha_i  \left\langle \phi_i, \phi_k' \right\rangle / M_{kk} \\
          &  \left\langle \rho^{-1} \tilde{\sigma}_{yz} , \phi_k' \right\rangle / M_{kk} = 
       \frac{\nu}{\rho} \sum_{i=0}^N \beta_i  \left\langle \phi_i, \phi_k' \right\rangle / M_{kk} \\
    \end{aligned}
\end{align}

and the terms $\left\langle \phi_i, \phi_k' \right\rangle =: D_{ik}$ are indeed symbolically integrable. 

The bulk friction part of the source term reads

\begin{python}
    def newtonian(self):
        out = Matrix([0 for i in range(self.n_fields)])
        offset = self.levels + 1
        h = self.variables[0]
        ha = self.variables[1 : 1 + self.levels + 1]
        hb = self.variables[1 + offset : 1 + self.levels + 1 + offset]
        p = self.parameters
        for k in range(1 + self.levels):
            for i in range(1 + self.levels):
                out[1 + k] += (
                    -p.nu
                    / h
                    * ha[i]
                    / h
                    * self.basismatrices.D[i, k]
                    / self.basismatrices.M[k, k]
                )
                out[1 + k + offset] += (
                    -p.nu
                    / h
                    * hb[i]
                    / h
                    * self.basismatrices.D[i, k]
                    / self.basismatrices.M[k, k]
                )
        return out
\end{python}

Combined, this results in the \code{source} term:

\begin{python}
    def source(self):
        out = Matrix([0 for i in range(self.n_fields)])
        out += Newtonian()
        out += slip_boundary_condition()
        return out
\end{python}

The code snippets above demonstrate that the hierarchical SME can be implemented and that the material closure can be easily exchanged with other material models. In particular, note that, unlike the SWE, the SME and other hierarchical methods have access to the vertical velocity profile. This enables the construction of closure relations based on gradients of the velocity, such as in a Newtonian fluid.

Figure \ref{fig:sme} shows the 3D reconstruction due to the implementation of \code{interpolate\_3d}. The test case was obtained using the SME with Legendre polynomials of order 4. The arrows in the image represent the vertical velocity profile at one particular position at the junction of the geometry. The elevation of the surface was obtained in a post-processing step.

\section{Conclusion}\label{sec:discussion}

In this article, we presented \textit{Zoomy}, a software framework for analyzing and simulating depth-averaged free-surface flow models.

Hierarchical models appear due to the description of the velocities and pressure $(u, v, w)$ and $p$ as polynomial expansions in the depth direction with a variable polynomial degree. The resulting depth-averaging and moment projections generate the model hierarchy.

\textit{Zoomy} addresses the research gap of systematically describing, analyzing, and solving hierarchical free-surface flow models generated by depth-averaging. 
In particular, users can
\begin{itemize}
    \item efficiently represent the hierarchical nature of the models,
    \item exchange the underlying basis functions,
    \item  perform the depth-integration symbolically,
    \item flexibly define the material closure and
    \item numerically solve the non-conservative PDE systems on 1D and 2D unstructured grids.
\end{itemize}
\textit{Zoomy} consists of a \textit{symbolic layer} to represent and analyze models. A code transformation creates a new representation of the model used by the \textit{numerical layer} to solve the PDE system numerically. These code transformations can have different output formats, for instance \code{Jax}, \code{NumPy} or \code{C} and allow a separation of the \textit{symbolic} and \textit{numerical layer}.

Based on four examples, we introduced the main building blocks needed to construct, analyze, and solve a model in \textit{Zoomy}. The examples highlighted how to

\begin{itemize}
    \item construct multi-dimensional models,
    \item represent model hierarchies,
    \item use the \textit{symbolic layer} for analysis and
    \item construct complex numerical solvers based on splitting schemes.
\end{itemize}
Current and future work includes extending the FVM-based numerical back-end, written in \code{Jax}, to the massively parallel block-structured adaptive-mesh-refinement framework \textit{AMReX}. Furthermore, we aim to create a more user-friendly interface for coupling the solver with the two-phase flow solver of \textit{OpenFOAM} with the coupling library \textit{preCICE} \cite{preCICEv2}.

\section*{Declarations}

\subsection*{Funding}
Ingo Steldermann was supported by the School for Data Science in Life, Earth, and Energy (HDS-LEE).

\subsection*{Code availability}
Zoomy is open-source software under the GNU General Public License v3.0 or later. 
The active source repository can be found at \url{https://github.com/mbd-rwth/Zoomy}.

This article refers to release version \code{1.0.0}, which is pinned as a corresponding release version in the source repository.

We make our project accessible by supporting installations via the \textit{conda} package manager \cite{anaconda_software_2016} or manual installation from source. Additionally, we support prebuilt container environments for Apptainer \cite{apptainer} and Docker \cite{merkel2014docker}.

\subsection*{Acknowledgements}
We thank Benjamin Terschanski for providing the excellent template of a GitHub repository and accompanying documentation, as well as his assistance in configuring the continuous integration pipeline and offering valuable comments on the manuscript draft.

\subsection*{Author contribution}
Ingo Steldermann: Writing - original draft, Visualization, Validation, Software, Methodology, Investigation, Formal analysis, Data curation, Conceptualization. \\ Julia Kowalski: Writing - review \& editing, Supervision, Resources, Funding acquisition, Conceptualization.

\subsection*{Use of editing tools and large language models}
The authors acknowledge the use of
Grammarly \url{https://grammarly.com} and Languagetool \url{https://languagetool.org/} for editing individual text passages.

\subsection*{Competing interests}
The authors declare no conflict of interest.

\bibliography{sn-bibliography}

\end{document}